\documentclass[eqsecnum,noshowpacs,showkeys,nofootinbib,aps]{revtex4}
\renewcommand{\theequation}{\arabic{equation}}
\def\beq{\begin{equation}}
\def\eeq{\end{equation}}
\def\bea{\begin{eqnarray}}
\def\eea{\end{eqnarray}}
\def\nn{\nonumber}
\def\pr{\prime}
\def\pa{\partial}

\def\na{\nabla}

\begin{document}
\title{Photon intrinsic frequency and size in stringy photon model}
\author{Soon-Tae Hong}
\email{galaxy.mass@gmail.com}
\affiliation{Center for Quantum Spacetime and Department of Physics, Sogang University, Seoul 04107, Korea}
\date{\today}

\begin{abstract}
Exploiting an open string which performs both rotational and pulsating motions, we investigate a photon intrinsic frequency. Explicitly evaluating the zero point fluctuation of the string which is delineated 
in terms of the quantum mechanical ground state energy in the vibrational mode of the string and the classical energy 
in the rotational mode, we find that the intrinsic frequency of the photon is given by $\omega_{\gamma}=9.00\times 10^{23}~{\rm sec}^{-1}$ and 
comparable to those of the baryons such as nucleon and delta baryon. Next, we calculate the photon size 
$\langle r^{2}\rangle^{1/2}(\rm photon)=0.17~{\rm fm}$ in a phenomenological stringy photon model. 
\end{abstract}
\maketitle

\section{Introduction}
\setcounter{equation}{0}
\renewcommand{\theequation}{\arabic{section}.\arabic{equation}}

It is well known that in the string theory a particle is assumed to be an extended object~\cite{green87,polchinski98}. In particular, in 
this theory the photon is given by the open string described in the extra dimensional space. Making use of the string version of 
the Hawking-Penrose singularity theorem~\cite{hawking70} and the corresponding stringy particle properties, 
we have studied~\cite{hong11,hong112} the stringy cosmology in a higher dimensional total spacetime to suggest that we can 
describe precisely the stringy congruence in terms of the universe expansion after the Big Bang. Note that in the stringy 
Hawking-Penrose singularity theorem, we have an advantage that in the early universe we have the degrees of freedom of the rotation and 
shear of stringy congruence. 

Next, the rotating string has been also investigated to obtain the propagator of a string in terms of 
a set of classical rigid rotators~\cite{sato}. Here the rotators have been assumed to possess some quantum excitations. 
Note that, in their approach, they have exploited the normal modes associated with the infinite sum over the product of 
creation and annihilation operators in quantum Hamiltonian. 

On the other hand, the hypersphere soliton model~\cite{man1,hongplb98,hong212} has been proposed to construct a topological lower bound on soliton energy and a set of equations of motion through the second class canonical 
quantization formalism~\cite{man1}. The baryon physical quantities such as baryon masses, charge radii and magnetic moments, have been later evaluated by making use of 
the canonical quantization in the hypersphere soliton model, to suggest that a realistic hadron physics can be described 
in terms of this phenomenological soliton~\cite{hongplb98}. Recently, we have calculated the intrinsic pulsating 
frequencies of the baryons~\cite{hong212}. To do this, we have used the hypersphere soliton model, 
where we have constructed the first class Hamiltonian to quantize the hypersphere soliton~\cite{man1,hongplb98,hong212}. In this first class Dirac quantization scheme, at first we have constructed the axial coupling constant in addition to the above physical quantities. 
These predictions~\cite{hong212} have been shown to be in good agreement with the corresponding experimental data. Next, we have evaluated 
explicitly the intrinsic frequencies $\omega_{N}$ and $\omega_{\Delta}$ of the nucleon and delta baryon, respectively. Moreover, we have found 
that the intrinsic frequency for more massive particle is greater than that for the less massive one. To be specific, we have constructed the identity $\omega_{\Delta}=2\omega_{N}$.

In this paper, we will propose a new phenomenological stringy photon model (SPM) to investigate the intrinsic frequency of pulsating photon, 
which is comparable to those~\cite{hong212} of the 
baryons such as nucleon and delta baryon. To do this, we will make use of the Nambu-Goto string 
theory~\cite{nambu70,goto71}, without resorting to the Ramanujan evaluation of the 
Riemann zeta function related with infinite mode sum~\cite{berndt85,green87,rosen00,tong,gova}. Next, in the SPM we will introduce an open string action associated with the photon~\cite{schwarz74}. Since the zero point energy is described in terms of the energies both for the quantum mechanical ground state and for 
the corresponding classical one~\cite{brink}, we will calculate the quantum mechanical ground state energy in the vibrational 
mode channel of the string and the classical energy in the rotational mode one. Making use of these two energies we will 
evaluate the zero point energy of the string to explicitly yield the photon intrinsic frequency. Here we will also find the formula for the 
the photon intrinsic frequency in terms of the photon mass, and then we will exploit the approximation that the photon mass is 
non-vanishing but negligible. Next, assuming that the photon size is given by the string radius in the SPM, we will 
predict the photon size by making use of the intrinsic frequency of pulsating photon. 

In Sec. II, we will briefly introduce basic features of an open string associated with a photon. In Sec. III, we will recapitulate the rotating 
string formalism. In Sec. IV, in the SPM we will explicitly construct the photon intrinsic frequency and the photon size. 
Next, we will discuss the chemical potential associated with the Bose-Einstein statistics for the massive photon with a finite size. Sec. V includes conclusions. In Appendix A, the Riemann zeta function is briefly discussed.

\section{Sketch of Nambu-Goto string theory}
\setcounter{equation}{0}
\renewcommand{\theequation}{\arabic{section}.\arabic{equation}}
\label{setupphotonsection}

In this section, before we construct the SPM, we will digress to pedagogically summarize a mathematical formalism for the 
Nambu-Goto open string which is related with a photon. 
In order to define the action on curved manifold, we introduce $(M,g_{ab})$ which is a 
$D$ dimensional spacetime manifold $M$ associated with the metric
$g_{ab}$. Given $g_{ab}$, we can have a unique covariant
derivative $\na_{a}$ satisfying~\cite{wald84}
\bea
\na_{a}g_{bc}&=&0,\nn\\
\na_{a}\omega^{b}&=&\pa_{a}\omega^{b}+\Gamma^{b}_{~ac}~\omega^{c},\nn\\
(\na_{a}\na_{b}-\na_{b}\na_{a})\omega_{c}&=&R_{abc}^{~~~d}~\omega_{d}.\label{rtensor}
\eea

We parameterize an open string by two world sheet coordinates
$\tau$ and $\sigma$, and then we have the corresponding vector
fields $\xi^{a}=(\pa/\pa\tau)^{a}$ and
$\zeta^{a}=(\pa/\pa\sigma)^{a}$.  The Nambu-Goto string action is
now given by~\cite{nambu70,goto71}
\beq 
S=-\kappa\int\int~d\tau
d\sigma f(\tau,\sigma),\label{nambugoto}
\eeq 
where the coordinates $\tau$ and
$\sigma$ have ranges $\tau_{1}\leq \tau\leq \tau_{2}$ and $0\leq \sigma\leq
\pi$ respectively and
\beq
f(\tau,\sigma)=[(\xi\cdot\zeta)^{2}-(\xi\cdot\xi)(\zeta\cdot\zeta)]^{1/2}.
\label{fts0}
\eeq
Here $\kappa$ is defined by $\kappa=\frac{1}{2\pi \alpha^{\prime}}$, with $\alpha^{\prime}$ being the slope of the Regge trajectories. 

We now perform an infinitesimal variation of the world sheets
$\gamma_{\alpha}(\tau,\sigma)$ traced by the open string during
its evolution in order to find the string geodesic equation from 
least action principle.  Here we impose the restriction that
the length of the string is $\tau$ independent. We introduce the deviation vector 
$\eta^{a}=(\pa/\pa \alpha)^{a}$ which represents the displacement to an infinitesimally
nearby world sheet, and we consider $\Sigma$ which denotes the three dimensional
submanifold spanned by the world sheets $\gamma_{\alpha}(\tau,\sigma)$.
We then may choose $\tau$, $\sigma$ and $\alpha$ as coordinates of
$\Sigma$ to yield the commutator relations 
\bea
\pounds_{\xi}\eta^{a}&=&\xi^{b}\na_{b}\eta^{a}-\eta^{b}\na_{b}\xi^{a}=0,\nn\\
\pounds_{\zeta}\eta^{a}&=&\zeta^{b}\na_{b}\eta^{a}-\eta^{b}\na_{b}\zeta^{a}=0,\nn\\
\pounds_{\xi}\zeta^{a}&=&\xi^{b}\na_{b}\zeta^{a}-\zeta^{b}\na_{b}\xi^{a}=0.\label{poundxizeta}
\eea 

Now we find the first variation as follows
\beq
\frac{dS}{d\alpha} =\int\int d\tau
d\sigma~\eta_{b}(\xi^{a}\na_{a}p^{b}+\zeta^{a}\na_{a}\pi^{b})
-\int d\sigma~p^{b}\eta_{b}|_{\tau=\tau_{1}}^{\tau=\tau_{2}}-\int
d\tau~\pi^{b}\eta_{b}|_{\sigma=0}^{\sigma=\pi},\label{dsdalpha2}
\eeq 
where the world sheet currents associated with $\tau$ and
$\sigma$ directions are respectively given by~\cite{scherk75}, 
\bea
p^{a}&=&\frac{\kappa}{f}[(\xi\cdot\zeta)\zeta^{a}-(\zeta\cdot\zeta)\xi^{a}],\nn\\
\pi^{a}&=&\frac{\kappa}{f}[(\xi\cdot\zeta)\xi^{a}-(\xi\cdot\xi)\zeta^{a}].
\label{pps2}\eea
Using the endpoint conditions 
\beq
\eta^{a}(\tau=\tau_{1};\sigma)=\eta^{a}(\tau=\tau_{2};\sigma)=0,
\eeq 
and 
\beq
\pi^{a}(\tau;\sigma=0)=\pi^{a}(\tau;\sigma=\pi)=0,
\label{end}
\eeq 
we have string geodesic equation
\beq
\xi^{a}\na_{a}p^{b}+\zeta^{a}\na_{a}\pi^{b}=0,\label{geodesicng}\eeq
and constraint identities~\cite{scherk75} 
\bea
p\cdot\zeta&=&0,~~~p\cdot p+\kappa^{2}\zeta\cdot\zeta=0,\nn\\
\pi\cdot\xi&=&0,~~~\pi\cdot \pi+\kappa^{2}\xi\cdot\xi=0.
\label{consts2} 
\eea

\section{Rotating open string in (3+1) dimensional spacetime}
\setcounter{equation}{0}
\renewcommand{\theequation}{\arabic{section}.\arabic{equation}}

In this section, to explicitly develop the SPM, we will evaluate the rotational energy of photon in the rotating open string 
theory defined on the $D=5$ dimensional total manifold, by restricting ourselves to exploiting one dimensional internal space 
parameterized by the coordinate $\sigma$, 
without appealing to the Ramanujan evaluation of the Riemann zeta function whose ambiguity is excluded in (\ref{noteq}) below. 
We thus consider the open string in the (3+1) dimensional flat 
spacetime and delineate the string in terms of the coordinates
\beq
x_{\mu}=(x_{0},x_{i})=(\tau,x_{i}(\tau;\sigma)),~~~(i=1,2,3).
\label{xmu123}
\eeq
We then find that $f(\tau,\sigma)$ in (\ref{fts0}) is given by
\beq
f(\tau,\sigma)=[(\dot{x}_{i}{x}^{\pr}_{i})^{2}+(1-\dot{x}_{i}\dot{x}_{i})x^{\pr}_{j}x^{\pr}_{j}]^{1/2}.
\label{fts}
\eeq
Here the overdot and prime denote derivatives with respect to $\tau$ and $\sigma$, respectively. In this paper, we 
use the metric signature $(+, -, -, -)$. Moreover we proceed to obtain the world sheet currents
\bea
p_{0}&=&\frac{\kappa}{f}x^{\pr}_{i}x^{\pr}_{i},~~~~~p_{i}=-\frac{\kappa}{f}[(\dot{x}_{j}x^{\pr}_{j})x^{\pr}_{i}-(x^{\pr}_{j}
x^{\pr}_{j})\dot{x}_{i}],\nn\\
\pi_{0}&=&-\frac{\kappa}{f}\dot{x}_{i}x^{\pr}_{i},~~~\pi_{i}=-\frac{\kappa}{f}[(\dot{x}_{j}x^{\pr}_{j})\dot{x}_{i}+(1-\dot{x}_{j}\dot{x}_{j})x^{\pr}_{i}].
\label{pppp}
\eea
Inserting $p_{\mu}$ and $\pi_{\mu}$ in (\ref{pppp}) into the string geodesic equation 
in (\ref{geodesicng}), one readily obtains
\bea
\frac{\pa}{\pa\tau}\left(\frac{x^{\pr}_{i}x^{\pr}_{i}}{f}\right)
-\frac{\pa}{\pa\sigma}\left(\frac{\dot{x}_{i}x^{\pr}_{i}}{f}\right)&=&0,\nn\\
\frac{\pa}{\pa\tau}\left[\frac{(\dot{x}_{j}x^{\pr}_{j})x^{\pr}_{i}-(x^{\pr}_{j}x^{\pr}_{j})\dot{x}_{i}}{f}\right]
+\frac{\pa}{\pa\sigma}\left[\frac{(\dot{x}_{j}x^{\pr}_{j})\dot{x}_{i}+(1-\dot{x}_{j}\dot{x}_{j})x^{\pr}_{i}}{f}\right]&=&0.
\label{eom2ns}
\eea
Exploiting the boundary conditions in (\ref{end}), one also finds at $\sigma=0$ and $\sigma=\pi$
\beq
\dot{x}_{i}x^{\pr}_{i}=0,~~~(1-\dot{x}_{j}\dot{x}_{j})x^{\pr}_{i}=0.
\label{end2}
\eeq

Next, in order to describe an open string, which is rotating in $(x_{1},x_{2})$ plane and residing 
on the string center of mass, we take an ansatz~\cite{sato}
\beq
x_{i}^{rot}=(r(\sigma)\cos\omega\tau,r(\sigma)\sin\omega\tau,0).
\label{xrot}
\eeq 
Here we propose that $r(\sigma)$ and $\omega$ represent respectively the diameter and angular velocity 
of the photon with solid spherical shape which is delineated by the open string. Note that $r(\sigma=\pi/2)$ denotes the center 
of the diameter of string. More specifically, $r(\sigma=\pi/2)$ is located in the center of the solid sphere 
which describes the photon. The first boundary condition 
in (\ref{end2}) is trivially satisfied and the second one yields
\beq
r^{\pr}(\sigma=0,\pi)=0.
\label{bc}
\eeq
We then obtain $r(\sigma)$ which fulfills the above condition in (\ref{bc}) 
\beq
r(\sigma)=\frac{1}{\omega}\cos\sigma.
\label{romega}
\eeq 
Note that the photon has a finite size which is filled with mass. 

Next, using the photon configuration in (\ref{xrot}) and (\ref{romega}) together with (\ref{pppp}), we readily find 
the rotational energy of the photon
\beq
E^{rot}=\int_{0}^{\pi}d\sigma~p_{0}^{rot}=\frac{1}{2\alpha^{\pr}\hbar\omega},
\label{hrot}
\eeq
where we have included $\hbar$ factor explicitly, and the value of $\alpha^{\pr}$ is given by~\cite{scherk75}
\beq
\alpha^{\pr}=0.95~{\rm GeV}^{-2}.\label{alphap}
\eeq
Note that the rotational degrees of freedom of the photon in the early universe have been investigated in Refs.~\cite{hong11,hong112}.

\section{Photon intrinsic frequency and photon size in SPM}
\setcounter{equation}{0}
\renewcommand{\theequation}{\arabic{section}.\arabic{equation}}
\label{photonsection}

In this section, we will explicitly evaluate the photon intrinsic frequency in the SPM. To do this, we 
calculate the vibrational energy of photon by introducing the string 
coordinate configurations
\beq
x_{i}=x_{i}^{rot}+y_{i},~~~i=1,2,3.
\label{yi}
\eeq 
Exploiting the coordinates in (\ref{yi}), we expand the string Lagrangian density
\beq
{\cal L}={\cal L}_{0}+\frac{1}{2}\frac{\pa^{2}{\cal L}}{\pa\dot{x}_{i}\pa\dot{x}_{j}}|_{0}\dot{y}_{i}\dot{y}_{j}
+\frac{\pa^{2}{\cal L}}{\pa\dot{x}_{i}\pa x^{\pr}_{j}}|_{0}\dot{y}_{i}y^{\pr}_{j}
+\frac{1}{2}\frac{\pa^{2}{\cal L}}{\pa x^{\pr}_{i}\pa x^{\pr}_{j}}|_{0}y^{\pr}_{i}y^{\pr}_{j}+\cdots,
\label{call}
\eeq
where the subscript $0$ denotes that the terms in (\ref{call}) are evaluated by using the coordinates 
in (\ref{xrot}). The ellipsis stands for the higher derivative terms. Here the first term is a constant given by 
${\cal L}_{0}={\cal L}(x_{i}^{rot})$. The first derivative terms vanish after exploiting the string geodesic equations 
(\ref{geodesicng}). Next in order to obtain the intrinsic vibration energy of photon, 
we define coordinates $z_{i}$ which co-rotates with the string itself
\bea
z_{1}&=&y_{1}\cos\omega\tau+y_{2}\sin\omega\tau,\nn\\
z_{2}&=&-y_{1}\sin\omega\tau+y_{2}\cos\omega\tau,\nn\\
z_{3}&=&y_{3}.
\label{zzz}
\eea  
After some straightforward algebra, one ends up with the Lagrangian density associated with the coordinates $z_{i}$
\bea
{\cal L}(z_{i})&=&\frac{\kappa}{2\sin^{2}\sigma}\left[\frac{1}{\omega}(\dot{z}_{2}+\omega z_{1})^{2}
+2\sin\sigma\cos\sigma((\dot{z}_{1}-\omega z_{2})z_{2}^{\pr}\right.\nn\\
&&\left.-(\dot{z}_{2}+\omega z_{1})z_{1}^{\pr})
-\omega z^{\pr 2}_{2}\right]+\frac{\kappa}{2\omega}(\dot{z}_{3}^{2}-\omega^{2}z_{3}^{\pr 2}).
\label{lagz}
\eea

The equations of motion for the directions $z_{2}$ and $z_{3}$ are given by
\bea
\ddot{z}_{2}+\omega^{2}z_{2}+2\omega^{2}\cot\sigma z_{2}^{\pr}-\omega^{2} z_{2}^{\pr\pr}&=&0,\nn\\
\ddot{z}_{3}-\omega^{2}z_{3}^{\pr\pr}&=&0.\label{z3eom}
\eea 
In order to construct the zero point energy $E^{zero}$ associated with the photon mass~\cite{brink}, we now investigate the quantum 
mechanical ground state energy of the string. Note that the photon is assumed to be in the ground state of the string energy spectrum. 
From (\ref{z3eom}) we find the eigenfunctions for the ground states
\bea
{z}_{2}&=&c_{2}\sin(\omega\tau+\phi_{2}),\label{z2sol}\\
{z}_{3}&=&c_{3}\cos\sigma\sin(\omega\tau+\phi_{3}).\label{z3sol}
\eea
Here $\phi_{2}$ and $\phi_{3}$ are arbitrary phase constants which are irrelevant to the physics arguments of interest. 

It seems appropriate to address comments on the photon vibration modes. The transverse mode $z_{2}$ in (\ref{z2sol}) 
is independent of the string coordinate $\sigma$, so that the photon can tremble back and forth with 
a constant amplitude, while the longitudinal mode $z_{3}$ in (\ref{z3sol}) possesses 
sinusoidal dependence on $\sigma$. Here note that $z_{3}$ does not move at the center of the string, namely at $\sigma=\pi/2$, 
independent of $\tau$ and the other parts of the string oscillate with the frequency $\omega$. As for the transverse mode $z_{1}$, 
one can readily find that any value for $z_{1}$ satisfies the Euler-Lagrange equation for $z_{1}$ obtained from 
the Lagrangian density in (\ref{lagz}). Up to now we have considered a single {\it massive} photon with the 
solid sphere shape, whose diameter is delineated in terms of the length of the open string. The photon thus has a disk-like cross section on which the coordinates $z_{1}$ and $z_{2}$ resides. Note that, similar to the phonon associated with massive particle lattice vibrations, the photon is massive so that we can have three polarization directions: two transverse directions as in the massless photon case, and 
an additional longitudinal one. Keeping this argument in mind, we find that there exist two transverse modes $z_{1}$ and $z_{2}$ associated with the photon vibrations on $z_{1}$-$z_{2}$ cross sectional disk, in addition to one longitudinal mode $z_{3}$. 
We thus have the transverse mode in $z_{1}$ direction to yield the eigenfunction for the ground state, 
with an arbitrary phase constant $\phi_{1}$ similar to $\phi_{2}$ and $\phi_{3}$ discussed above,
\beq
{z}_{1}=c_{1}\sin(\omega\tau+\phi_{1}).\label{z1sol}
\eeq
Note that, as in the case of massless photon, $z_{1}$ mode oscillates with the same frequency $\omega$ as $z_{2}$ mode does.

Note also that the above solutions $z_{i}$ satisfy their endpoint conditions at $\sigma=0$ and $\sigma=\pi$
\beq
z_{i}^{\pr}=0,~~~i=1,2,3.
\eeq
The energy eigenvalues in the ground states in (\ref{z2sol})--(\ref{z1sol}) are then given by
\beq
E^{vib}_{i}=\frac{1}{2}\hbar \omega,~~~i=1,2,3.\label{evib123}
\eeq
Exploiting the energies in (\ref{evib123}), we arrive at the vibrational energy of the open string ground state
\beq
E^{vib}=\sum_{i=1}^{3}E^{vib}_{i}=\frac{3}{2}\hbar\omega.\label{evibtot}
\eeq

The zero point energy $E^{zero}$ of the string is known to be the difference between the energy for the quantum mechanical ground state and the corresponding classical one~\cite{brink}. Note that, in the SPM the energy for the quantum mechanical ground state is given by 
$E^{vib}$ in (\ref{evibtot}), while the classical 
energy is given by $E^{rot}$ in (\ref{hrot}). Moreover, the zero point energy state describes the photon mass 
$M_{\gamma}$ to produce $E^{zero}=M_{\gamma}$ at $\omega=\omega_{\gamma}$, so that we can find the photon mass of the form~\cite{brink} 
\beq
M_{\gamma}=E^{vib}_{\gamma}-E^{rot}_{\gamma}.\label{propose}
\eeq
Note that we have already removed the translational degree of freedom, by considering the string observer residing on the photon center of mass. 
Substituting $E^{rot}_{\gamma}$ in (\ref{hrot}) and $E^{vib}_{\gamma}$ in (\ref{evibtot}) in the case of $\omega=\omega_{\gamma}$ into (\ref{propose}), we find 
\beq
M_{\gamma}=\frac{3}{2}\hbar\omega_{\gamma}-\frac{1}{2\alpha^{\pr}\hbar\omega_{\gamma}},\label{propose2}
\eeq
to yield 
\beq
\omega_{\gamma}=\frac{1}{3\hbar}\left[M_{\gamma}+\left(M_{\gamma}^{2}+\frac{3}{\alpha^{\pr}}\right)^{1/2}\right].\label{propose3}
\eeq
On the other hand, one has the upper limit of experimental value for the photon mass~\cite{pdg}
\beq
M_{\gamma}^{exp}=1.00\times 10^{-27}~{\rm GeV}.\label{mexp}
\eeq
Since in (\ref{propose3}) the photon mass $M_{\gamma}$ is non-vanishing but negligible, compared with $1/\alpha^{\pr}$ term, we obtain 
\beq
\omega_{\gamma}\cong \frac{1}{\hbar(3\alpha^{\pr})^{1/2}}.\label{omegapred}
\eeq
Inserting the experimental data for $\hbar$ and $\alpha^{\pr}$ in (\ref{alphap}) into (\ref{omegapred}), we are finally left with 
\beq
\omega_{\gamma}=9.00\times 10^{23}~{\rm sec}^{-1}.
\label{omegagamm}
\eeq
As shown in Table~\ref{tableintrinsicfre}, the predicted value of the intrinsic frequency for the bosonic photon 
with spin one is comparable to those for the fermionic nucleon and delta baryon with spin 1/2~\cite{hong212}. 
To be more specific, the intrinsic frequency with spin one particle is greater than that with spin 1/2 particle. Note that 
$\omega_{\gamma}$ is interpreted both as the angular velocity of the rotating solid spherical shape photon in 
(\ref{xrot}), and as the intrinsic frequency associated with the vibrational modes in (\ref{omegagamm}).

Next, making use of (\ref{romega}) and (\ref{omegagamm}), we predict the photon size given by the string radius in the SPM
\beq
\langle r^{2}\rangle^{1/2}(\rm photon)=\frac{c}{2\omega_{\gamma}}=0.17~{\rm fm},
\label{photonradius}
\eeq
where we have explicitly included the speed of light $c$. The above value of $\langle r^{2}\rangle^{1/2}(\rm photon)$ is 
21\% of the proton magnetic charge radius $\langle r^{2}\rangle^{1/2}_{M}(\rm proton)=0.80~{\rm fm}$~\cite{hongplb98,hong212}.

\begin{table}[t]
\caption{The intrinsic frequencies of particles.}
\begin{center}
\begin{tabular}{lll}
\hline
Particle type  &Notation &Intrinsic frequency\\
\hline
Nucleon &$\omega_{N}$ &$0.87\times 10^{23}~{\rm sec}^{-1}$\\
Delta baryon &$\omega_{\Delta}$ &$1.74\times 10^{23}~{\rm sec}^{-1}$\\
Photon &$\omega_{\gamma}$ &$9.00\times 10^{23}~{\rm sec}^{-1}$\\
\hline
\end{tabular}
\end{center}
\label{tableintrinsicfre}
\end{table}

Now, we have a couple of comments to address. First, we investigate the thermodynamic properties of the 
{\it massive photon with finite size} which are described in terms of 
the Bose-Einstein statistics~\cite{reif}, where the total number of the photons is fixed by 
the constraint $\sum_{r}n_{r}=N$. Here $n_{r}$ and $N$ are the number of photons in quantum state $r$ and 
the total photon number, respectively. In this case, we exploit 
the Bose-Einstein distribution $\bar{n}_{s}=\frac{1}{e^{\beta(\epsilon_{s}-\mu)}-1}$, where $\beta=1/kT$ with $k$ 
and $T$ are the Boltzmann constant and temperature, respectively, and $\epsilon_{s}$ $(s=1,2,...)$ is 
the energy in state $s$. Here $\mu$ is the chemical potential per the massive photon defined by $\mu=\frac{\pa F}{\pa N}$, 
with $F$ being the Helmholtz free energy. Note that, in order for the Bose-Einstein distribution to be reduced 
to the Planck distribution for the massless photon statistics, one should have the vanishing chemical 
potential which means that we need infinitely many photons. 

Second, note that the photon mass creation is possible in the Big Bang epoch~\cite{hong11,hong112}. Moreover the equation of state 
$\frac{D-4}{D-2}\rho+\frac{D}{D-2}P\ge 0$ derived in the string cosmology~\cite{hong11} yields $\rho+5P\geq 0$ for the
 massive extended (not point) particles in the $D=5$ dimensional total manifold associated with the SPM. Note also that, 
in the standard cosmology for the point massless particles, the equation of state is described by $\rho+P\geq 0$. 
Now we propose that all the particles including the photons in the universe are massive so that we can have only a single 
equation of state $\rho+5P\geq 0$.

\section{Conclusions}
\setcounter{equation}{0}
\renewcommand{\theequation}{\arabic{section}.\arabic{equation}}

In summary, we have proposed a new approach of the SPM to investigate a 
photon intrinsic frequency by exploiting a string which performs both rotational 
and pulsating motions. Here we have interpreted the string as a diameter of a solid spherical shape photon. 
Explicitly we have found that the intrinsic frequency of the photon is comparable to those of the baryons such as 
nucleon and delta baryon. Next, we have discussed the chemical 
potential associated with the Bose-Einstein statistics for the massive photon with a finite size. 
Moreover, in the SPM we have evaluated the photon size given by the string radius which is approximately 21\% of 
the proton magnetic charge radius. It will be interesting to search for a strong experimental evidence for the photon size which 
could be associated with the manifest photon phenomenology such as the photoelectric effect, Compton scattering and 
Raman scattering, for instance. Once this is done, the algorithm for the stringy photon could give 
some progressive impacts on the realistic precision optics. Assuming that the SPM exploited in this paper could 
be a precise description for the photon, the newly evaluated photon intrinsic frequency 
$\omega_{\gamma}=9.00\times 10^{23}~{\rm sec}^{-1}$ and photon size $\langle r^{2}\rangle^{1/2}(\rm photon)=0.17~{\rm fm}$ 
could be fundamental predictions in the extended object phenomenology, similar to $\omega_{N}$, $\omega_{\Delta}$ and the charge radii 
in the hypersphere soliton model~\cite{hong212,man1,hongplb98}. 

\acknowledgments{The author was supported by Basic Science Research Program through 
the National Research Foundation of Korea funded by the Ministry of Education, NRF-2019R1I1A1A01058449.}

\appendix
\section{Ramanujan evaluation for Riemann zeta function}\label{zetachapter}
\setcounter{equation}{0}
\renewcommand{\theequation}{A.\arabic{equation}}

The Riemann zeta function has been developed in theoretical physics in addition to pure mathematics. 
In this Appendix, we will revisit the Ramanujan result for evaluation the Riemann zeta function 
which is associated with the (super)string theory defined on the 
$D=26$ ($D=10$) dimensional total manifold. Note that, in the light-cone gauge quantization in the bosonic 
string theory, the anomaly associated with the 
commutator of the Lorentz generators has been canceled in the $D=26$ critical dimensions and with the condition 
\beq
a=-\frac{D-2}{2}\zeta(-1)=1,
\eeq 
where the Ramanujan evaluation for the Riemann zeta function $\zeta(-1)=\sum_{n=1}^{\infty}n=-\frac{1}{12}$ 
has been used~\cite{green87,gova}. After evaluating the Riemann zeta function following the Ramanujan scheme~\cite{berndt85,rosen00} 
which is a pure number theoretical approach, in this Appendix we will proceed to point out the origin of ambiguity involved in the scheme to clarify the uniqueness of the Riemann zeta function. 

\subsection{Evaluation of the Riemann zeta function in the Ramanujan scheme}
\label{Riemann zeta function1}

We will now heuristically investigate the Riemann zeta function, by following the Ramanujan scheme~\cite{berndt85} to add up positive integers 
associated with mode sum in the string theory. To do this, we start with the Riemann zeta function defined 
as~\cite{berndt85,green87,gova,rosen00,tong,broda}
\beq
\zeta(s)=\sum_{n=1}^{\infty}\frac{1}{n^{s}}=1+\frac{1}{2^{s}}+\frac{1}{3^{s}}+\cdots.
\eeq
Now we are interested in the particular value of the Riemann zeta function
\beq
\zeta(-1)=\sum_{n=1}^{\infty}\frac{1}{n^{-1}}=1+2+3+\cdots.
\label{zeta}
\eeq
To calculate the value of $\zeta(-1)$ defined above, by following the Ramanujan scheme~\cite{berndt85}, 
we consider the identity
\beq
\frac{1}{1-x}=1+x+x^{2}+x^{3}+\cdots
\eeq
to yield
\beq
\frac{1}{(1-x)^{2}}=1+2x+3x^{2}+4x^{3}+\cdots.
\label{1x2}
\eeq
In the case of $x=-1$, (\ref{1x2}) yields
\beq
1-2+3-4+\cdots=\frac{1}{4}.
\label{14}
\eeq

On the other hand, manipulating the Riemann zeta function $\zeta(-1)$ in (\ref{zeta}) produces~\cite{berndt85}
\beq
\zeta(-1)-4\zeta(-1)=-3\zeta(-1)=1-2+3-4+\cdots.
\label{zeta14}
\eeq
Combining (\ref{14}) with (\ref{zeta14}) yields~\cite{berndt85}
\beq
\zeta(-1)\equiv 1+2+3+\cdots=-\frac{1}{12},
\label{zetaminus12}
\eeq
which has been applied to the string theory associated with the critical dimensions $D=26$. 

Next we consider the infinite sum
\beq
\sum_{r=1,3,5,\cdots}^{\infty}\frac{r}{2}=\frac{1}{2}+\frac{3}{2}+\frac{5}{2}+\cdots=\frac{1}{2}\zeta(-1)-\zeta(-1)=\frac{1}{24},
\label{sum135}
\eeq
which has been also exploited in the $D=10$ superstring theory. Note that the 
Ramanujan result of the Riemann zeta function $\zeta(-1)$ is derived through the number theoretical approach. 
In other words, the derivation of (\ref{zetaminus12}) is purely algebraic and it has nothing to do with the regularization scheme related with 
the Casimir force~\cite{tong} which will be discussed below. Moreover, the Ramanujan result in (\ref{zetaminus12}) has been 
derived without resorting to the analytic continuation procedure~\cite{green87}. Note that, if we follow the Ramanujan result for evaluation 
of the Riemann zeta function, we have an ambiguity that $\zeta(-1)=-\frac{1}{12}$ in addition to the well known result $\zeta(-1)=\infty$.

\subsection{Uniqueness of the Riemann zeta function}
\label{Riemann zeta function2}

Next, we will scrutinize the Riemann zeta function, to 
clarify the ambiguity associated with the Ramanujan 
result in (\ref{zetaminus12}). To do this, we consider the infinity $\infty$ which satisfies the basic algebra
\bea
\infty\pm\alpha&=&\infty,\label{inftypma}\\
\alpha\times\infty&=&\infty,\label{alphainfty}\\
\infty-\alpha\times \infty&&{\rm is~indeterminate},\label{inftyminus}
\eea
where $\alpha$ is a finite positive real number. In the case of $\alpha=4$ in (\ref{inftyminus}), we arrive at
\beq
\infty-4\times \infty~~{\rm is~indeterminate}.\label{1m4}
\label{infty14}
\eeq
Now (\ref{1m4}) can be readily checked if we consider the explicit relations obtained from (\ref{alphainfty}), for instance
\beq
4\times \infty=3\times \infty=2\times \infty=\infty.
\eeq

Note that the left hand side of (\ref{zeta14}) is the same as $\infty-4\times \infty$, and thus we 
conclude that, due to (\ref{infty14}), we cannot uniquely obtain the Ramanujan 
result in (\ref{zeta14}) to yield
\beq
\zeta(-1)-4\zeta(-1)\neq -3\zeta(-1).
\label{zetawrong}
\eeq
In other words, we end up with the statement
\beq
\zeta(-1)-4\zeta(-1)~{\rm is~indeterminate}.
\label{zetanone}
\eeq
We thus cannot proceed to evaluate $\zeta(-1)-4\zeta(-1)$ at this stage, and the ensuing calculations leading to 
(\ref{zetaminus12}) and (\ref{sum135}) are incorrect to yield
\bea
\zeta(-1)&=&1+2+3+\cdots\neq-\frac{1}{12},\nn\\
\sum_{r=1,3,5,\cdots}^{\infty}\frac{r}{2}&=&\frac{1}{2}+\frac{3}{2}+\frac{5}{2}+\cdots\neq\frac{1}{24}.
\label{noteq}
\eea
Since the Ramanujan approach to the Riemann zeta function is now excluded, we then have the uniqueness of 
the Riemann zeta function
\beq
\zeta(-1)\equiv 1+2+3+\cdots=\infty,
\label{uniquezeta}
\eeq
without any ambiguity. In other words, we have only one desired unique value $\infty$ for the infinite series $1+2+3+\cdots$.

Parenthetically, we make comments on the finite sum
\beq
\sum_{n=1}^{N}\frac{1}{n^{-1}}=1+2+3+\cdots+N=\frac{N(N+1)}{2}.
\eeq
In this case, we arrive at
\beq
\sum_{n=1}^{N}\frac{1}{n^{-1}}-4\sum_{n=1}^{N}\frac{1}{n^{-1}}=-\frac{3N(N+1)}{2}.
\eeq
The incorrectness in $\zeta(-1)-4\zeta(-1)=-3\zeta(-1)$ involved in (\ref{zeta14}) seems to originate from 
neglecting the difference between the infinite sum and finite sum. For the correct statement, see (\ref{zetawrong}).

Next, we have comments on the calculation of the Riemann zeta function through the regularization scheme~\cite{tong}.
Even though the Ramanujan derivation of (\ref{zetaminus12}) is purely algebraic, we assume that the regularization scheme {\it could} be applicable to evaluation of the Riemann zeta function, and then let us see what happens later. To do this, we start with the regularization procedure of the Riemann zeta function $\zeta(-1)$ by replacing the divergent sum over integers by the expression~\cite{tong}\footnote{In Ref.~\cite{tong}, there exists a typo in the expression in third line of (\ref{regul}).}
\bea
\sum_{n=1}^{\infty}n=1+2+3+\cdots&\rightarrow& \lim_{\epsilon\rightarrow 0}\sum_{n=1}^{\infty}ne^{-\epsilon n}\nn\\
&=&\lim_{\epsilon\rightarrow 0}\left(-\frac{\pa}{\pa\epsilon}\right)\sum_{n=1}^{\infty}e^{-\epsilon n}\nn\\
&=&\lim_{\epsilon\rightarrow 0}\left(-\frac{\pa}{\pa\epsilon}\right)(e^{\epsilon}-1)^{-1}\nn\\
&=&\lim_{\epsilon\rightarrow 0}\left(\frac{1}{\epsilon^{2}}-\frac{1}{12}+{\cal O}(\epsilon)\right)\nn\\
&=&\infty-\frac{1}{12}\nn\\
&=&\infty.
\label{regul}
\eea
In the fifth line of (\ref{regul}), we have used the simple mathematical relation 
$\lim_{\epsilon\rightarrow 0}\frac{1}{\epsilon^{2}}=\infty$, and in the last line we have exploited the identity (\ref{inftypma}). 
We are thus again left with the uniqueness of the Riemann zeta 
function $\zeta(-1)$ in (\ref{uniquezeta}) even in the case that we exploit the regularization procedure. 

Note that in Ref.~\cite{tong} the term $\frac{1}{\epsilon^{2}}$ in (\ref{regul}) has been renormalized away. However there 
is no reason to discard this divergent term, in mathematical and physical view points, since we need to just add up positive integers associated with mode sum as in the case of Ramanujan scheme, to evaluate the Riemann zeta function. If we {\it were} in the 
shoes of Tong~\cite{tong} who exploits the renormalization-away method, we could also manipulate the result in (\ref{regul}) as follows
\bea
\lim_{\epsilon\rightarrow 0}\sum_{n=1}^{\infty}ne^{-\epsilon n}&=&\lim_{\epsilon\rightarrow 0}
\left(\frac{1}{\epsilon^{2}}-\frac{1}{12}-p+{\cal O}(\epsilon)+p\right)\nn\\
&=&\lim_{\bar{\epsilon}\rightarrow 0}
\left(\frac{1}{\bar{\epsilon}^{2}}+p+{\cal O}(\epsilon)\right),
\eea
where we have introduced the definition of $\bar{\epsilon}$: $\frac{1}{\bar{\epsilon}^{2}}=\frac{1}{\epsilon^{2}}-\frac{1}{12}-p$ 
for a given finite number $p$. In this case, after renormalizing away the term $\frac{1}{\bar{\epsilon}^{2}}$, we {\it could} find the Riemann zeta function $\zeta(-1)=p$ for any given number $p$, to yield again the ambiguity in fixing the value of the Riemann zeta function, similar to the ambiguity in Ramanujan scheme associated 
with (\ref{zetanone}). Note that the Ramanujan scheme has been employed~\cite{berndt85} 
to calculate the Riemann zeta function $\zeta(-1)$ {\it without} resorting to the renormalization-away method.

\end{document}